# Protecting entanglement by adjusting the velocities of moving qubits inside non-Markovian environments

Ali Mortezapour[1]*, Mahdi Ahmadi Borji[1], Rosario Lo Franco[2,3]

[1]Department of Physics, University of Guilan, P. O. Box 41335–19141, Rasht, Iran
[2]Dipartimento di Energia, Ingegneria dell'Informazione e Modelli Matematici, Università di Palermo, Viale delle Scienze, Edificio 9, 90128 Palermo, Italy
[3]Dipartimento di Fisica e Chimica, Università di Palermo, via Archirafi 36, 90123, Palermo, Italy
Corresponding author E-mail: mortezapour@guilan.ac.ir

**Abstract**. Efficient entanglement preservation in open quantum systems is a crucial scope towards a reliable exploitation of quantum resources. We address this issue by studying how two-qubit entanglement dynamically behaves when two atom qubits move inside two separated identical cavities. The moving qubits independently interact with their respective cavity. As a main general result, we find that under resonant qubit-cavity interaction the initial entanglement between two moving qubits remains closer to its initial value as time passes compared to the case of stationary qubits. In particular, we show that the initial entanglement can be strongly protected from decay by suitably adjusting the velocities of the qubits according to the non-Markovian features of the cavities. Our results supply a further way of preserving quantum correlations against noise with a natural implementation in cavity-QED scenarios and are straightforwardly extendable to many qubits for scalability.

## I. Introduction
Quantum mechanics is a fundamental and accurate theory that can predict many striking effects that are counterintuitive and impossible in classical mechanics. One of the most remarkable effects is the existence of entangled states of two or more distant particles.

Entanglement, as non-local quantum coherence of a composite system, describes quantum correlations between two or more quantum subsystems. Generally, a multipartite system is said to be entangled, within a mathematical language, if its quantum state cannot be written as the tensor product of the quantum states of the constituent subsystems and is thus a whole object. Nowadays, entanglement is of great importance due to its fundamental role in quantum information tasks such as quantum teleportation [1], quantum error correction [2, 3], quantum cryptography [4] and quantum dense coding [5]. However, the unavoidable interaction between any realistic quantum system and its surrounding environment typically causes decoherence and then usually destroys the entanglement of a quantum state. Such an undesired effect has been considered as a pesky problem to the implementation of various quantum information processing schemes. Therefore, finding a way to overcome this drawback is a valuable step for the realization of practical quantum computers [6, 7].

Along this route, investigation of entanglement dynamics in Markovian and non-Markovian environments has recently attracted wide attention [8-64]. It is well known that in Markovian (memoryless) environments the single qubit coherence decays exponentially. As a consequence, two-qubit entanglement will deteriorate in a short time, also completely disappearing at a finite time [65, 66]. Nevertheless, researchers have recently succeeded to preserve two-qubit entanglement in Markovian environments [8-12]. For instance, it has been shown that two-qubit entanglement can be protected by quantum interferences [8]. Moreover, two-qubit entanglement can be both generated and kept for a very long time in the ideal left-handed materials [9].

On the other hand, compared to Markovian environments, non-Markovian (memory-keeping) environments such as structured cavities or photonic band gap (PBG) mediums show distinctly different effects on decoherence and entanglement [13-33]. Under this condition, it was shown that the entanglement can revive after a finite time of complete disappearance (dark period) [34, 35]. The mechanism of such entanglement behavior has been explored in Refs. [36-39]. Furthermore, some recent investigations have reported that two-qubit entanglement in non-Markovian environments can be protected by detuning [40-42], quantum Zeno effect [43], dynamical decoupling pulse sequences [44-48], compound quantum environments [49-53], suitable classical noise [54-63] and continuous driving field [64].

However, so far the effect of translational motion of qubits on the two-qubit entanglement dynamics in non-Markovian environment has not been addressed. The knowledge of this effect is important in that it can provide novel insights for preserving entanglement against noise by tailoring motional properties of the qubits. This boosts our motivation to address this problem by studying a simple yet paradigmatic system made of two moving qubits independently interacting with their local environments. It is assumed that the environment of each qubit is structured and modeled by a leaky cavity with Lorentzian spectral density and the qubits are initially prepared in a maximally entangled state.

The paper is organized as follows. In Sec. II the Hamiltonian is introduced and a state evolution in the considered system is discussed. In Sec. III concurrence is presented as measure of entanglement. In Sec. IV, we present the results of our numerical simulations illustrating the excellent performance of qubit motion in protecting two-qubit entanglement. Finally, Sec. V concludes this paper.

**II. Model and Solution**

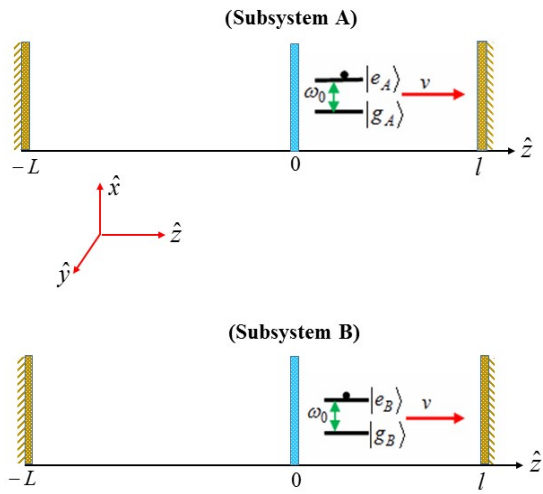

**Figure 1.**

The system under consideration is composed of two non-interacting identical subsystems (subsystem A and subsystem B), each one consisting of an atom qubit and a structured environment made of two perfect reflecting mirrors at the positions $z = -L$ and $z = l$ with a partially reflecting mirror in $z = 0$. This creates a sort of two consecutive cavities $(-L, 0)$ and $(0, l)$ as shown in figure 1. Any classical electromagnetic field in each subsystem $(-L, l)$ may be expanded in terms of the exact monochromatic modes $U_k(z)$ at frequency $\omega_k = ck$ [67-69]:

$$E(z,t) = \sum_k E_k(t) U_k(z) + c.c, \qquad (1)$$

where $E_k(t)$ is the amplitude in the $k$ th-mode. In order to satisfy the boundary conditions at the mirrors, the mode functions $U_k(z)$ can only be of the following form:

$$U_k(z) = \begin{cases} \xi_k \sin k(z+L), & z < 0 \\ M_k \sin k(z-l), & z > 0 \end{cases}, \qquad (2)$$

Here $\xi_k$ has the values 1, -1 going from each mode to the subsequent one, and $M_k$ for a good cavity ($r \approx 1$) has the expression

$$M_k = \frac{(c\lambda^2/l)^{1/2}}{[(\omega_k - \omega_n)^2 + \lambda^2]^{1/2}}, \qquad (3)$$

where $\omega_n = n\pi c/l$ ($n \gg 1$) are the frequencies of the quasi modes and $\lambda$ is the damping of the ($0, l$) cavity. In fact, $\lambda$ quantifies photons leakage through cavity mirrors and indicates the spectral width of the coupling.

The qubit (two-level atom) is taken to interact only with the second cavity ($0, l$) and it moves along the z-axis with constant velocity $v$ (see figure 1). This condition can be thought to be fulfilled by Stark shifting (for instance, by turning on a suitable external electric field) the atom frequency far out of resonance from the cavity modes until $z = 0$ and then turning off the Stark shift [71]. During the translational motion, the qubits interact with their respective cavity modes. The Hamiltonian of each subsystem ($i = A, B$) under the dipole and rotating-wave approximation is given by (we take $\hbar = 1$ and omit index $i$)

$$H = \omega_0 |e\rangle\langle e| + \sum_k \omega_k a_k^\dagger a_k + \sum_k M_k f_k(z)[g_k |e\rangle\langle g| a_k + g_k^* |g\rangle\langle e| a_k^\dagger], \qquad (4)$$

where $|e\rangle$ ($|g\rangle$) and $\omega_0$ are the previously mentioned excited (ground) state and the transition frequency of the qubit, $a_k^\dagger$ ($a_k$) is the creation (annihilation) operator for the $k$-th cavity mode with frequency $\omega_k$ and $g_k = -d(\omega_k/\hbar\varepsilon_0 Al)^{1/2}$ denotes the coupling constant between the qubit and the cavity modes. Note that, $d$ is the magnitude of electric-dipole moment of the atom qubit and $A$ is the surface area of the cavity mirrors.

The parameter $f_k(z)$ describes the shape function of qubit motion along the z-axis, and it is given by [69]

$$f_k(z) = f_k(vt) = \sin[k(z-l)] = \sin[\omega_k(\beta t - \tau)], \qquad (5)$$

where $\beta = v/c$ and $\tau = l/c$. Note that the coupling function is not zero when $z = 0$, while it is zero when $z = l$ (perfect boundary). It is noteworthy that the qubits can be realized by the circular $^{85}Rb$ Rydberg atoms with the two circular levels with principal quantum numbers 51 and 50 called $|e\rangle$ and $|g\rangle$ respectively. For such a qubit, the transition frequency is at $\omega_0 = 51.1\,\text{GHz}$, corresponding to the decay rate $\gamma = 33.3\,\text{Hz}$ [70, 71].

It is well known that the translational motion of an atom can be treated classically if the de Broglie wavelength $\lambda_B$ of the atom is much smaller than the wavelength $\lambda_0$ of the resonant transition [69, 72]. For the considered qubit, this is equivalent to

$$\lambda_B/\lambda \approx 10^{-19}\beta^{-1} \ll 1. \qquad (6)$$

On the other hand, the relative smallness of photon momentum ($\hbar\omega_0/c$) compared to atomic momentum ($mv$) allows one to neglect the atomic recoil resulting from the interaction with the electric field [73]. To satisfy this condition, the velocity of the qubit is required to be $v > 10^{-7}$ m/s.

Owing to the two subsystems are the same and noninteracting, they experience independent evolutions. Therefore, one can analyze only the dynamics of one subsystem and then use it to obtain the evolution of the total system [14, 34, 35]. The evolution is considered during the flight time of the atom qubit inside the cavity. We assume each subsystem to be initially in a product state with the qubit in a coherent superposition of its states ($C_e(0)|e\rangle + C_g(0)|g\rangle$) and the cavity modes in the vacuum state ($|0\rangle$),

$$|\Psi(0)\rangle = \{C_e(0)|e\rangle + C_g(0)|g\rangle\}|0\rangle. \tag{7}$$

Hence, at any later time $t$, the quantum state of a subsystem can be written as

$$|\Psi(t)\rangle = C_e(t)|e\rangle|0\rangle + C_g(0)|g\rangle|0\rangle + \sum_k C_{g,k}(t)|g\rangle|1_k\rangle, \tag{8}$$

in which the cavity field state $|1_k\rangle$ describes the presence of a single photon in mode $k$, and $C_{g,k}(t)$ represents its probability amplitude. Using the time-dependent Schrödinger equation, the differential equations for the probability amplitudes $C_e(t)$ and $C_{g,k}(t)$ are given by

$$i\dot{C}_e(t) = \omega_0 C_e(t) + \sum_k g_k M_k f_k(vt) C_{g,k}(t), \tag{9}$$

$$i\dot{C}_{g,k}(t) = \omega_k C_{g,k}(t) + g_k^* M_k f_k(vt) C_e(t). \tag{10}$$

Solving Eq. (10) formally and substituting the solution into Eq. (9), one obtains

$$\dot{C}_e(t) + i\omega_0 C_e(t) = -\int_0^t dt' C_e(t') \sum_k |g_k|^2 M_k^2 f_k(vt) f_k(vt') e^{-i\omega_k(t-t')}, \tag{11}$$

By redefining the probability amplitude as $C_e(t') = \tilde{C}_e(t') e^{-i\omega_0 t'}$, we can rewrite Eq. (11) as

$$\dot{\tilde{C}}_e(t) + \int_0^t dt' F(t,t') \tilde{C}_e(t') = 0., \tag{12}$$

where the kernel $F(t,t')$, which is the correlation function including the memory effect, is

$$F(t,t') = \sum_k |g_k|^2 M_k^2 f_k(vt) f_k(vt') e^{-i(\omega_k-\omega_0)(t-t')}. \tag{13}$$

In the continuous limit the kernel above becomes

$$F(t,t') = \int_0^\infty J(\omega_k) \sin[\omega_k(\beta t - \tau)] \sin[\omega_k(\beta t' - \tau)] e^{-i(\omega_k-\omega_0)(t-t')} d\omega_k, \tag{14}$$

where $J(\omega_k)$ is the spectral density of an electromagnetic field inside the $(0,l)$ cavity and has the form [74]

$$J(\omega_k) = \frac{1}{2\pi} \frac{\gamma \lambda^2}{[(\omega_0 - \omega_k - \Delta)^2 + \lambda^2]}, \quad (15)$$

where $\Delta = \omega_0 - \omega_n$ is the detuning between $\omega_0$ and the center frequency of the cavity modes ($\omega_n$). $\gamma = (d^2 \omega_n / \hbar \varepsilon_0 A l)$ is the decay rate of the qubit in the Markovian limit of flat spectrum with the qubit is at rest in a position in which $U_k(z)$ has a maximum. The relaxation time scale $\tau_q$ over which the state of the system changes is then related to $\gamma$ by $\tau_q \approx \gamma^{-1}$. As noted above, the parameter $\lambda$ indicates the spectral width of the coupling, and it is related to the cavity correlation time $\tau_{cav}$ via $\tau_{cav} = \lambda^{-1}$. The weak and strong coupling regimes can be distinguished by comparing $\tau_{cav}$ and $\tau_q$. The weak coupling regime corresponds to the case $\tau_q > 2\tau_{cav}$ ($\gamma < \lambda/2$) while the opposite relationship $\tau_q < 2\tau_{cav}$ (or $\gamma > \lambda/2$) holds in the strong coupling regime where non-Markovian effects become relevant [14, 34].

Substituting the resulting kernel into Eq. (8) yields

$$\dot{\tilde{C}}_e(t) + \frac{\lambda^2 \gamma}{2\pi} \int_0^\infty \left\{ \frac{\sin[\omega_k(\beta t - \tau)]}{[(\omega_0 - \omega_k - \Delta)^2 + \lambda^2]} \int_0^t dt' \sin[\omega_k(\beta t' - \tau)] e^{-i(\omega_k - \omega_0)(t-t')} \tilde{C}_e(t') \right\} d\omega_k = 0. \quad (16)$$

By calculating $C_e(t)$, the reduced density matrix of each qubit $\rho_s(t)$ can be written in the basis $\{|e\rangle, |g\rangle\}$ as

$$\rho_s(t) = \begin{pmatrix} |C_e(t)|^2 & C_g^*(0) C_e(t) \\ C_g(0) C_e^*(t) & 1 - |C_e(t)|^2 \end{pmatrix}. \quad (17)$$

Taking the derivative of Eq. (17) with respect to time, we get

$$\dot{\rho}_s(t) = -i \frac{\Omega(t)}{2} [\sigma_+ \sigma_-, \rho_s(t)] + \frac{\Gamma(t)}{2} [2\sigma_- \rho_s(t) \sigma_+ \\ - \sigma_+ \sigma_- \rho_s(t) - \rho_s(t) \sigma_+ \sigma_-], \quad (18)$$

where $\Omega(t) = -2 \operatorname{Im}\left[\frac{\dot{C}_e(t)}{C_e(t)}\right]$ and $\Gamma(t) = -2 \operatorname{Re}\left[\frac{\dot{C}_e(t)}{C_e(t)}\right]$. The quantity $\Omega(t)$ plays the role of a time-dependent Lamb shift and $\Gamma(t)$ can be interpreted as a time-dependent decay rate [74]. Once known the dynamics of individual subsystems, we can now determine the dynamics of the total system. In the standard product basis $\{|1\rangle \equiv |e_A e_B\rangle, |2\rangle \equiv |e_A g_B\rangle, |3\rangle \equiv |g_A e_B\rangle,$ $|4\rangle \equiv |g_A g_B\rangle\}$ and assuming that the two individual subsystems are the same, one obtains the elements of the time-dependent reduced density matrix of the total system $\rho(t)$ as [34]

$$\rho_{11}(t) = \rho_{11}(0)|C_e(t)|^4$$
$$\rho_{22}(t) = \rho_{22}(0)|C_e(t)|^2 + \rho_{11}(0)|C_e(t)|^2[1-|C_e(t)|^2]$$
$$\rho_{33}(t) = \rho_{33}(0)|C_e(t)|^2 + \rho_{11}(0)|C_e(t)|^2[1-|C_e(t)|^2]$$
$$\rho_{44}(t) = 1 - \rho_{11}(t) - \rho_{22}(t) - \rho_{33}(t),$$
$$\rho_{12}(t) = \rho_{12}(0)|C_e(t)|^2 C_e(t),$$
$$\rho_{13}(t) = \rho_{13}(0)|C_e(t)|^2 C_e(t), \quad (19)$$
$$\rho_{14}(t) = \rho_{14}(0)C_e(t)^2,$$
$$\rho_{23}(t) = \rho_{23}(0)|C_e(t)|^2,$$
$$\rho_{24}(t) = \rho_{24}(0)C_e(t) + \rho_{13}(0)C_e(t)[1-|C_e(t)|^2],$$
$$\rho_{34}(t) = \rho_{34}(0)C_e(t) + \rho_{12}(0)C_e(t)[1-|C_e(t)|^2].$$

with $\rho_{ji}^*(t) = \rho_{ij}(t)$.

### III. Concurrence

The entanglement of a 2 × 2 quantum system, described by the density matrix $\rho$, can be measured by concurrence which is defined as [75]

$$C(\rho) = \max\{0, \sqrt{\lambda_1} - \sqrt{\lambda_2} - \sqrt{\lambda_3} - \sqrt{\lambda_4}\}, \quad (20)$$

the quantities $\lambda_i$ being the eigenvalues of the matrix $\tilde{\rho} = \rho(\sigma_y^A \otimes \sigma_y^B)\rho^*(\sigma_y^A \otimes \sigma_y^B)$ arranged in decreasing order of magnitude. Here $\rho^*$ is the complex conjugate of $\rho$ in the standard basis, and $\sigma_y$ denotes the Pauli matrix. The concurrence varies from $C=0$ for a disentangled state to $C=1$ for a maximally entangled state. We will restrict our study to the following initial Bell (maximally entangled) state

$$|\Psi\rangle = (|e_A e_B\rangle + |g_A g_B\rangle)/\sqrt{2}. \quad (21)$$

The corresponding concurrence at time $t$ is given by [34]

$$C_\Psi(\rho) = \max\{0, 2|\rho_{14}(t)| - 2\sqrt{\rho_{22}(t)\rho_{33}(t)}\} \quad (22)$$

### IV. Results

In this section, we analyze the dynamics of two-qubit entanglement by numerical solution of Eq. (16) via the fourth-order Runge-Kutta method and show how the translational motion of qubits can protect the entanglement. Hereafter, the parameter $\beta\omega_0$ is exploited to represent the information about the velocity of the qubits. Moreover, all the parameters are scaled by $\gamma$ chosen as the unit.

Figure 2 illustrates the time behavior of the concurrences $C_\Psi$ for various velocities of the qubits. It is assumed that $\lambda = 0.01\gamma$ and the qubits resonantly interact with their respective cavities ($\Delta = 0$).

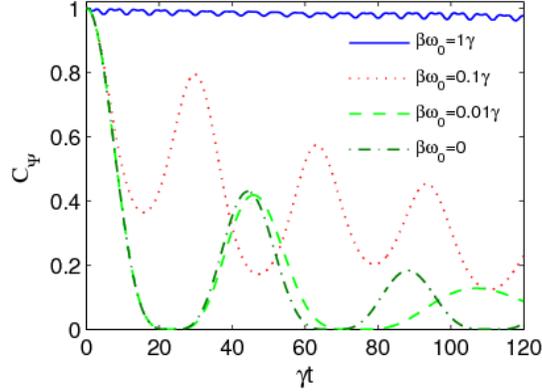

**Figure 2.**

It is seen that in the cases of stationary ($\beta\omega_0 = 0$) and slowly moving ($\beta\omega_0 = 0.01$) qubits, the entanglement appears to damp out or collapse and, after a short time, it revives. At longer times, one can find a sequence of collapses and revivals. The peaks of the revivals decrease as time increases and the entanglement eventually vanishes after a number of fluctuations. When $\beta\omega_0 = 0.1\gamma$ the entanglement reduces with an oscillatory behavior, but it does not disappear at a finite time. Nevertheless, entanglement once again tends to be zero in the long-time limit. An interesting result is obtained for $\beta\omega_0 = 1\gamma$. In this case, the entanglement is strongly protected from being lost. Namely, the entanglement does not change appreciably from its initial value as time passes.

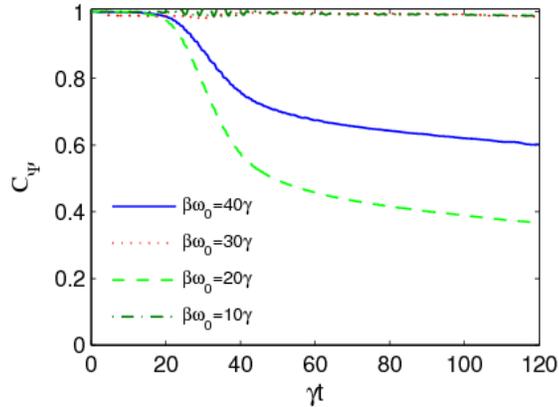

**Figure 3.**

According to figure 2 one may conclude that a stronger entanglement protection can be obtained by regularly increasing the velocity of the qubits. To deepen this conjecture, we examine the effect of higher velocities on the entanglement dynamics in figure 3. The atom-cavity parameters are the same as in figure 2 except the atom velocities. It is shown that the entanglement is strongly protected for $\beta\omega_0 = 10\gamma$ and $\beta\omega_0 = 30\gamma$ but, when $\beta\omega_0 = 20\gamma$ and $\beta\omega_0 = 40\gamma$, entanglement perceptibly decreases after about $\gamma t = 20$ with the decay for $\beta\omega_0 = 20\gamma$ being faster than for $\beta\omega_0 = 40\gamma$. Hence, this suggests that higher velocities do not necessarily guarantee stronger and better entanglement protection. Furthermore, it seems that the degree of entanglement exhibits a sort of periodic behavior by increasing the velocities of the qubits. To corroborate this observation, we plot concurrence $C_\Psi$ versus different velocities of the qubits for $\gamma t = 100$ in figure 4. In particular, we display and compare the values of the $C_\Psi$ for $\beta\omega_0 = (n)5\gamma$ where $n = 0,1,2,..$ at the dimensionless time $\gamma t = 100$. The data are specified with stars and connected by dotted line to show the global behavior. It is seen that the entanglement shows minima at $\beta\omega_0 = (2m)10\gamma$ and maxima at $\beta\omega_0 = (2m-1)10\gamma$ ($m = 1,2...$). We find out a contrast between trends of maxima and minima by increasing $m$: the value of the minima is significantly increased by increasing $m$, whereas the value of the maxima is slightly decreased.

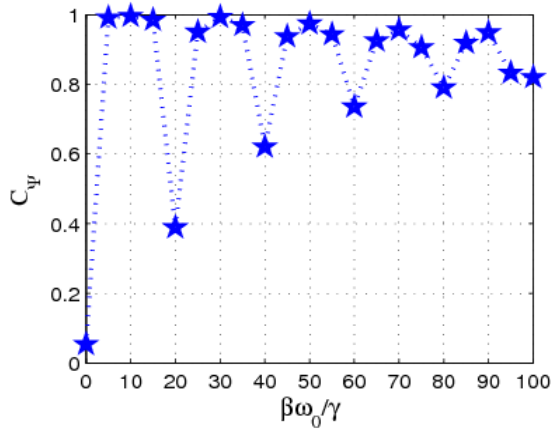

**Figure 4.**

To probe the phyisical origin behind the aforementioned entanglement behavior, we plot in figure 5 the decay rate of single qubit ($\Gamma(t)/\gamma$) as a function of $\gamma t$ for different velocities of the qubit. As can be seen from this figure, the decay rate can be greatly inhibited for suitable values of qubit velocity.

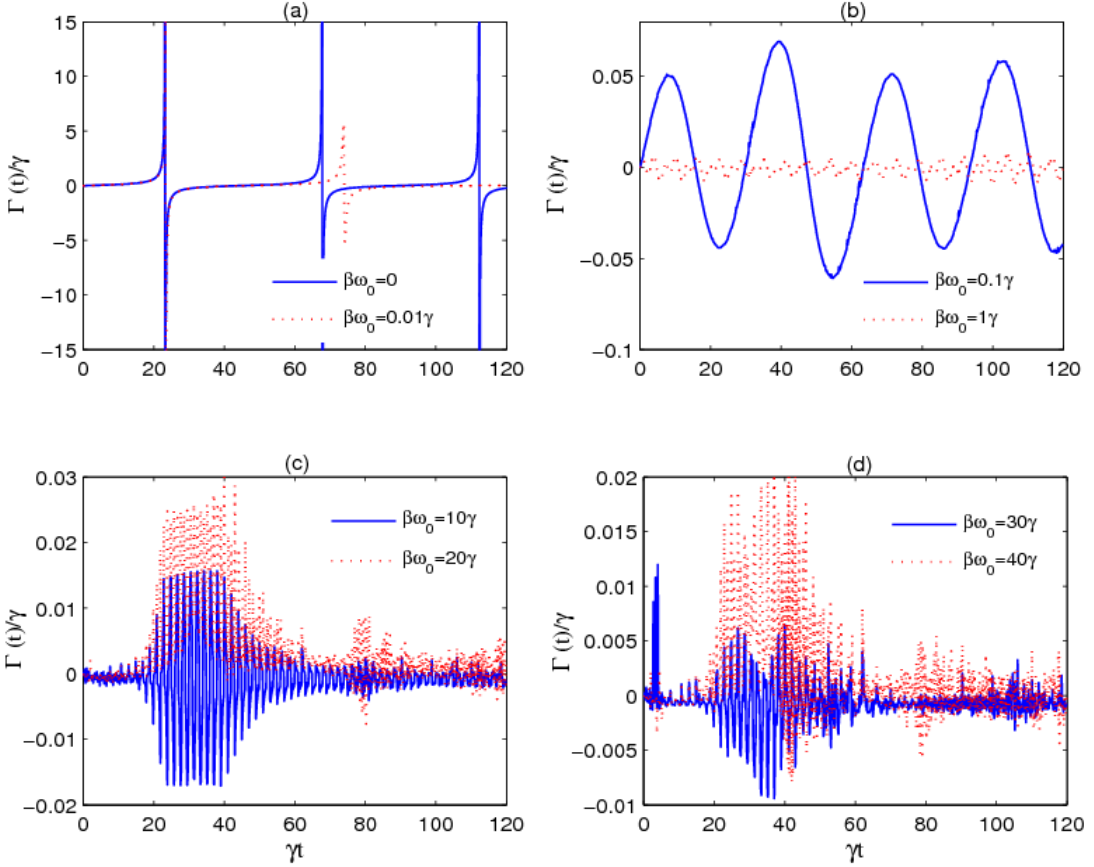

**Figure 5.**

It is known that the negativeness of the decay rate indicates reabsorbtion of emitted photon by the qubit, leading to information backflow from the environment to the qubit and thus signaling non-Markovianity [13-15, 30]. Figure 5(a) shows that the decay rate of stationary qubit ($\beta\omega_0 = 0$) has a periodic behavior with positive and negative spikes in it. The periodicity is increased for $\beta\omega_0 = 0.01\gamma$, whereas the magnitude of the spikes is decreased. This change in the evolution of the decay rate explains why the second revival of entanglement for $\beta\omega_0 = 0.01\gamma$ occurs later than $\beta\omega_0 = 0$ (see figure 2). In figure 5(b), we see that the spikes are completely disappeared and the decay rate gets oscillating behavior. However, the amplitude and periodicity of the decay rate are remarkably decreased by incresing $\beta\omega_0$ from $0.01\gamma$ to $1\gamma$. In fact, such a decrease in amplitude and periodicity respectively leads to the strong entanglement protection and its fast oscillation for $\beta\omega_0 = 1\gamma$ displayed in figure 2. In figure 5(c) and (d), we observe that the decay rate exhibits fast fluctuations. Note that the decay rate for $\beta\omega_0 = 20\gamma$ and $\beta\omega_0 = 40\gamma$ gets almost only positive values after $\gamma t \geq 20$. This fact means very weak non-Markovian conditions and justifies the decaying behavior of entanglement in figure 3.

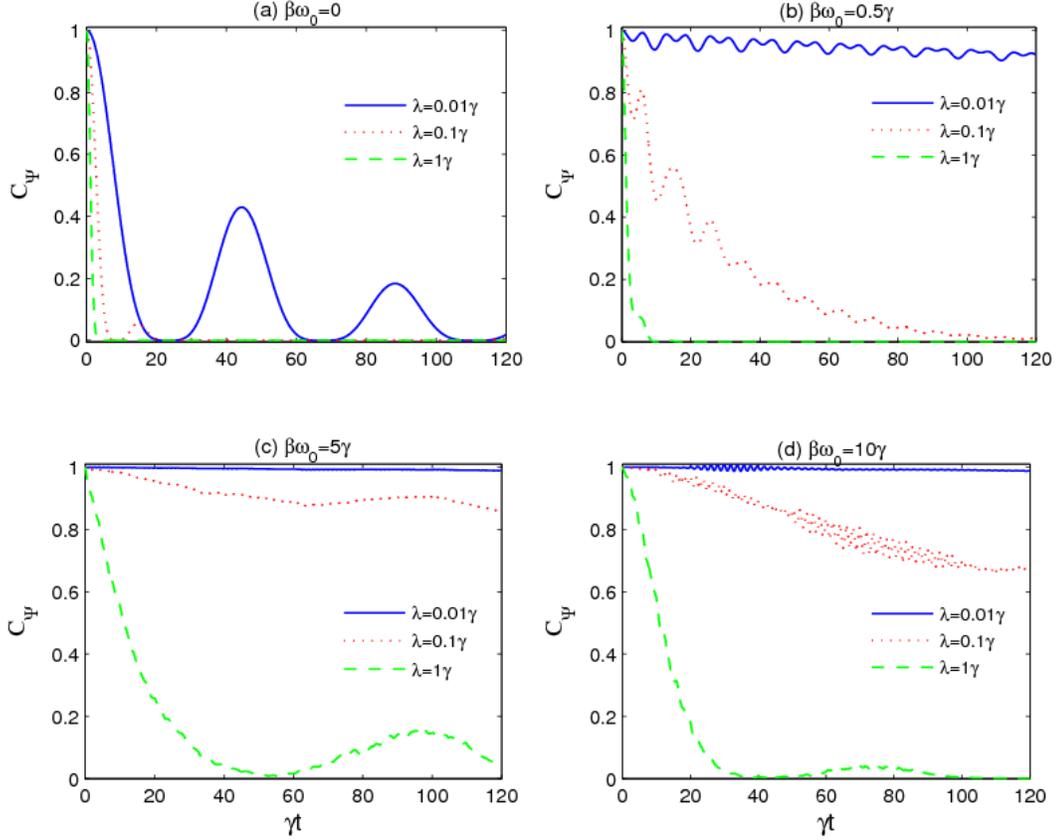

**Figure 6.**

Figure 6 displays the effect of $\lambda$ on the time evolution of concurrence $C_\Psi$ for various velocities of the qubits under the resonant condition. Tracking the behavior of the entanglement in the panels of figure 6 reveals that, for a given velocity, the entanglement can be maintained for a longer time by increasing the non-Markoviniaty (memory effects) of the system dynamics (that is, decreasing the cavity bandwidth $\lambda$). Furthermore, figure 6 shows that the motion of the qubits with high enough velocities can remarkably affect the entanglemet even in environments with weaker memory effects (larger values of the cavity bandwidth $\lambda$). Another fact is that, when $\lambda = 0.1\gamma$ or $\lambda = 1\gamma$, entanglement protection for $\beta\omega_0 = 5\gamma$ is better than for $\beta\omega_0 = 10\gamma$ (compare figure 6(c) and (d)). It entirely confirms the result of figure 3: increasing velocities do not always ensure a more efficient preservation of entanglement. However, all these results under resonant interaction indicate that, compared to the case of stationary qubits, quantum entanglement between moving qubits generally remains nearer to its initial value as time goes by.

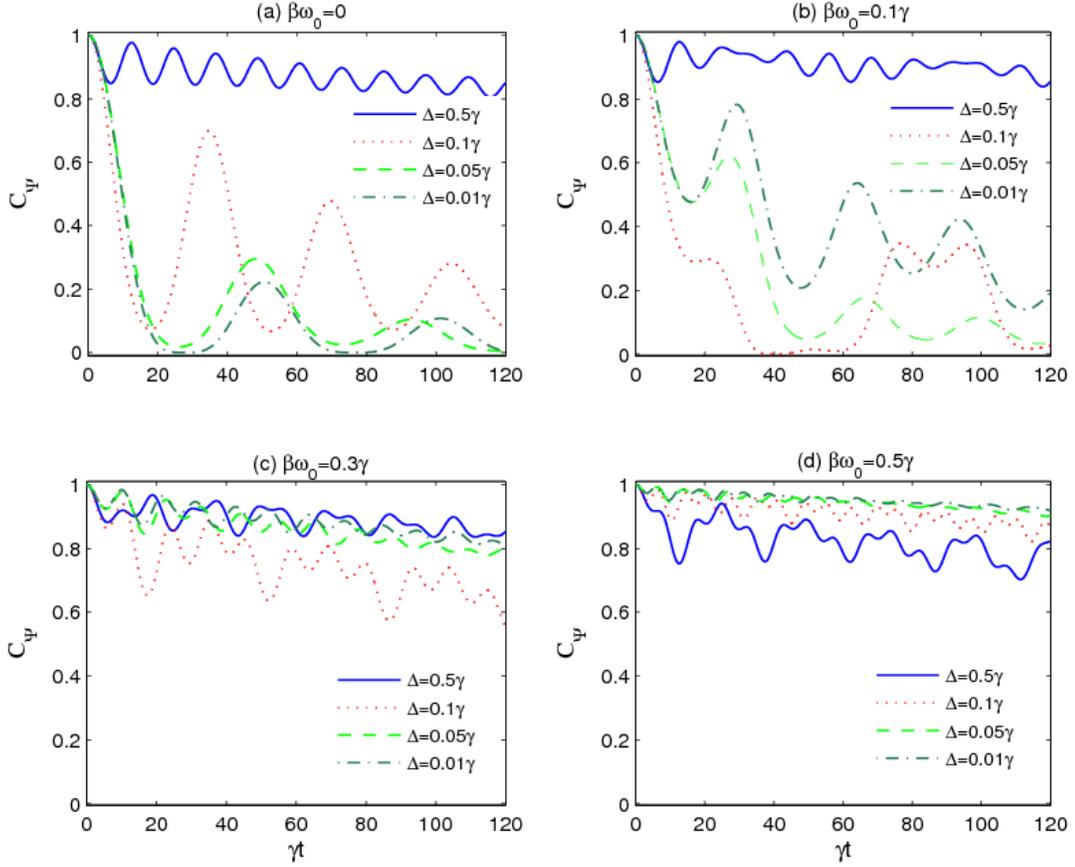

**Figure 7.**

Now we turn our attention to the off-resonant ($\Delta \neq 0$) qubit-cavity interactions. In figure 7(a–d), we display the effect of detuning on the time evolution of the concurrence ($C_\Psi$) for various velocities of the qubits when $\lambda = 0.01\gamma$. It can be clearly seen that in the case of stationary qubits, the entanglement increases regularly by gradual growth of the detuning parameter (Figure 7(a)). A similar result was reported before (see Refs. [41, 42]). However, such a regular behavior is not observed for moving qubits (Figure 7(b)-(d)). Indeed, motion of the qubits disturbs the growth of entanglement by increasing $\Delta$. However, our calculations show that, when $\Delta \geq 1\gamma$, a strong entanglement protection can be achieved for both stationary and moving qubits, with the case of moving qubits being more effective. Based on all the reported results (displayed from figure 2 to figure 7), we can conclude that entanglement can be efficiently protected by suitably adjusting the velocities of the moving qubits according to the characteristics of the cavities.

It is worth to mention that the value of the parameter $\beta\omega_0 = (x)\gamma$ is equivalent to a velocity $v = 0.2(x)\,\text{m/s}$ for the considered atom qubits, where $x$ is a real coefficient. This entails that the range of velocities considered in this work is $2 \times 10^{-3}$ m/s $\leq v \leq 20$ m/s. Notice that such a range of velocity values largely fulfills Eq. (3) and permits us to neglect atomic recoil.

Recent significant progress in QED experiments allows the achievement of the cavity parameters used in this work. For instance, ultrahigh finesse Fabry-Perot superconducting resonant cavities with quality factors $Q \approx 4.2 \times 10^{10}$, corresponding to the spectral width $\lambda \approx 7$ Hz ($\tau_{cav} = 130$ ms) at $\omega_0 = 51.1$ GHz, have been realized [76]. In addition, nowadays circuit quantum electrodynamics (circuit-QED) technologies are also capable to create a position dependent qubit-cavity coupling strength with sinusoidal function (like that of Eq. (2)) [77, 78]: adjusting the position of the qubit linearly with time, so to have a relation like $z = vt$, will be then sufficient to realize the model here considered.

**Conclusion**

We have introduced a system containing two moving atom qubits inside two separated identical cavities. The moving qubits independently interact with their respective cavity modes. In the case of resonant interaction, we have found that compared to stationary qubits, quantum entanglement between moving qubits generally remains closer to its initial value over time. Moreover, we have demonstrated that the initial entanglement can be strongly protected for a long time by appropriately setting the velocities of the qubits according to the non-Markovian characteristics of the cavities (e.g., quality factor, spectral bandwidth and mode frequencies). A strong-coupling (low spectral bandwidth) is in general required for a more effective entanglement protection, with an interplay between memory effects and decoherence inhibition indicated by the value of the qubit decay rate. The role of the translational movement of the atoms in controlling the decay rate stems from the presence of atom (qubit) velocities in the intensity of the coupling to the cavities (see equations (4) and (5)). These considerations suggest a quantitative study, to be done elsewhere, of the degree of non-Markovianity by suitable geometric measures [79] in order to assess to what extent memory effects contribute to preserve entanglement for various velocities of the qubits.

Our results supply new knowledge regarding the response of the entanglement dynamics to the velocity of an atomic qubit interacting with a cavity radiation, which suggests a further way of preserving quantum correlations against noise with a natural implementation in cavity-QED scenarios. Thanks to the analogy between cavity-QED and circuit-QED, one can then design feasible circuit-QED schemes which reproduce our model, by exploiting currently available setups with a sinusoidal position-dependent qubit-cavity coupling [77, 78] where the qubit position slowly varies linearly with time.

Our model is straightforwardly extendable to a set of many independent qubits, thus satisfying the requirement of scalability. The present study opens the way to investigate other situations of interest, for instance an array of consecutive cavities where each atom qubit can fly in or a set of oscillating atoms each one within its own cavity, which may lead to new efficient strategies for entanglement protection.

**Figure captions**

**Figure 1.** Schematic illustration of a setup in which two qubits are moving inside two distinct but identical cavities. The two qubits are initially entangled, but afterwards they have no direct interaction.

**Figure 2.** Concurrence $C_\Psi$ as a function of scaled time $\gamma t$ for various velocities of the qubits: $\beta\omega_0 = 0$ (dash-dotted dark green line), $\beta\omega_0 = 0.01\gamma$ (dashed light green line), $\beta\omega_0 = 0.1\gamma$ (dotted red line), $\beta\omega_0 = 1\gamma$ (solid blue line). The values of the other parameters are taken as follows: $\Delta = 0$, $\lambda = 0.01\gamma$.

**Figure 3.** Concurrence $C_\Psi$ as a function of scaled time $\gamma t$ for various velocities of the qubits; $\beta\omega_0 = 10\gamma$ (dash-dotted dark green line), $\beta\omega_0 = 20\gamma$ (dashed light green line), $\beta\omega_0 = 30\gamma$ (dotted red line), $\beta\omega_0 = 40\gamma$ (solid blue line). Other parameters are the same as those used in figure 2.

**Figure 4.** Concurrence $C_\Psi$ as a function of velocity of the qubits ($\beta\omega_0 = 5(n)\gamma$ where $n = 0,1,2,..$) at $\gamma t = 100$ for $\lambda = 0.01\gamma$ and $\Delta = 0$. The stars represent the data points connected by dotted lines.

**Figure 5.** The decay rate of single qubit $\Gamma(t)$ as a function of scaled time $\gamma t$ for various velocities of the qubit (a) $\beta\omega_0 = 0$ and $\beta\omega_0 = 0.01\gamma$ (b) $\beta\omega_0 = 0.1\gamma$ and $\beta\omega_0 = 1\gamma$ (c) $\beta\omega_0 = 10\gamma$ and $\beta\omega_0 = 20\gamma$ (d) $\beta\omega_0 = 30\gamma$ and $\beta\omega_0 = 40\gamma$. Other parameters are the same as those used in figure 2.

**Figure 6.** Concurrence $C_\Psi$ as a function of scaled time $\gamma t$ for various velocities of the qubits; (a) $\beta\omega_0 = 0$, (b) $\beta\omega_0 = 0.5\gamma$, (c) $\beta\omega_0 = 5\gamma$, (d) $\beta\omega_0 = 10\gamma$ in different non-Markovian environments: $\lambda = 0.01\gamma$ (solid-blue line), $\lambda = 0.1\gamma$ (dotted-red line), $\lambda = 1\gamma$ (dashed light green line), $\Delta = 0$.

**Figure 7.** Concurrence $C_\Psi$ as a function of scaled time $\gamma t$ for various velocities of the qubits; (a) $\beta\omega_0 = 0$, (b) $\beta\omega_0 = 0.1\gamma$, (c) $\beta\omega_0 = 0.3\gamma$, (d) $\beta\omega_0 = 0.5\gamma$ and for various values of detuning: $\Delta = 0.01\gamma$ (Dashed dark green line), $\Delta = 0.05\gamma$ (Dashed light green line), $\Delta = 0.1\gamma$ (Dotted red line), $\Delta = 0.5\gamma$ (Solid blue line), $\lambda = 0.01\gamma$.